\documentclass[%
 aip,
 rsi,
 amsmath,amssymb,
 reprint,%
]{revtex4-1}

\usepackage{graphicx}
\usepackage{dcolumn}
\usepackage{bm}

\usepackage[utf8]{inputenc}
\usepackage[T1]{fontenc}
\usepackage{mathptmx}
\usepackage{graphicx}
\usepackage{amssymb, amsmath, enumerate, wasysym}
\usepackage{cancel}
\usepackage{mathtools}

\newcommand{\um}{{\mu\mathrm{m}}}
\renewcommand{\vec}{\mathbf}

\begin{document}

\title{Cryogen-free variable temperature scanning SQUID microscope}
\author{Logan Bishop-Van Horn}
\altaffiliation[Present address: ]{Quantum Circuits, Inc., 25 Science Park Suite 203, New Haven, CT 06511, USA.}
\affiliation{Stanford Institute for Materials and Energy Sciences, SLAC National Accelerator Laboratory, 2575 Sand Hill Road, Menlo Park, CA 94025, USA.}
\affiliation{Department of Physics, Stanford University, Stanford, CA 94305, USA.}

\author{Zheng Cui}
\affiliation{Stanford Institute for Materials and Energy Sciences, SLAC National Accelerator Laboratory, 2575 Sand Hill Road, Menlo Park, CA 94025, USA.}
\affiliation{Department of Applied Physics, Stanford University, Stanford, CA 94305, USA.}

\author{John R. Kirtley}
\affiliation{Geballe Laboratory for Advanced Materials, Stanford University, Stanford, CA 94305, USA.}

\author{Kathryn A. Moler}
\altaffiliation[Email address: ]{kmoler@stanford.edu}
\affiliation{Stanford Institute for Materials and Energy Sciences, SLAC National Accelerator Laboratory, 2575 Sand Hill Road, Menlo Park, CA 94025, USA.}
\affiliation{Department of Applied Physics, Stanford University, Stanford, CA 94305, USA.}
\affiliation{Geballe Laboratory for Advanced Materials, Stanford University, Stanford, CA 94305, USA.}

\date{\today}
%
%
\begin{abstract}
Scanning Superconducting QUantum Interference Device (SQUID) microscopy is a powerful tool for imaging local magnetic properties of materials and devices, but it requires a low-vibration cryogenic environment, traditionally achieved by thermal contact with a bath of liquid helium or the mixing chamber of a “wet” dilution refrigerator. We mount a SQUID microscope on the 3 K plate of a Bluefors cryocooler and characterize its vibration spectrum by measuring SQUID noise in a region of sharp flux gradient. By implementing passive vibration isolation, we reduce relative sensor-sample vibrations to 20 nm in-plane and 15 nm out-of-plane. A variable-temperature sample stage that is thermally isolated from the SQUID sensor enables measurement at sample temperatures from 2.8 K to 110 K. We demonstrate these advances by imaging inhomogeneous diamagnetic susceptibility and vortex pinning in optimally-doped YBCO above 90 K. 
\end{abstract}

\maketitle

%
%
\section{Introduction}
Scanning Superconducting QUantum Interference Device (SQUID) microscopy has developed over the past 20+ years into a valuable component of the experimental toolbox for fundamental and applied condensed matter physics, enabling advances in the understanding of cuprate and pnictide high-$T_\text{c}$ superconductors (Ref.~\onlinecite{kirtley_reppp2010} and references therein), quantum effects in mesoscopic normal metal\cite{persistent_prl2009} and superconducting\cite{fluxoid_prb2011,fluctuation_science2007} rings, superconductivity and magnetism in complex oxide heterostructures,\cite{laosto_nphys2011, laosto_prb2012,laosto_nano2012,laosto_ncomm2012} edge currents in topological\cite{edge_nmat2013,edge_prl2014} and trivial\cite{edge_njp2016} phases, and current-phase relations in exotic Josephson junctions.\cite{cpr_nano2013,cpr_prl2015,cpr_nphys2017}

As a scanning probe technique with a superconducting sensor, scanning SQUID microscopy requires a low-vibration cryogenic environment,
traditionally achieved by placing the microscope in thermal contact with the liquid helium bath or 1 K pot of a ${}^4$He cryostat, or the mixing chamber of a ``wet'' dilution refrigerator pre-cooled with liquid ${}^4$He. With rising helium prices and increasingly frequent helium supply interruptions in the U.S., many research institutions are turning to cryogen-free refrigerators such as pulse tube cryocoolers to reach temperatures of $\sim4\;\mathrm{K}$ without the use of liquid ${}^4$He.

Cryogen-free systems eliminate loss of ${}^4$He and disruptions to experiments due to liquid ${}^4$He transfers, and can provide a larger cryogenic experimental volume than wet systems. However, since cooling in a pulse tube cryocooler is achieved by pulsing high pressure helium gas through a pulse tube at a frequency of roughly 1.4 Hz,\cite{pt_cryogenics1996} low frequency (1 Hz - 1 kHz) vibrations are a concern in cryogen-free systems,\cite{vib_cryogenics2010,vib_nuclear2017,vib_rsi2006} particularly for scanning probe measurements.\cite{dgg_rsi2013,vib_rsi2014,vib_arxiv2018}

The sensitivity of a scanning SQUID measurement to relative sensor-sample motion caused by external vibrations depends on the sample being measured, the mode of operation of the SQUID sensor, and size of the SQUID pickup loop and/or field coil. However, given the sub-micron spatial resolution of our current generation of SQUID sensors,\cite{sssm_rsi2016} our scanning SQUID microscopes can generally tolerate relative sensor-sample motion of order 10 nm.

With this in mind, we built a scanning SQUID microscope in a 3 K pulse tube cryocooler, characterized the sensor-sample vibration spectrum, and implemented simple, low-cost passive vibration isolation to reduce the sensor-sample motion to an acceptable level. We also thermally isolated the microscope from its sample mount, allowing us to measure samples at temperatures up to 110 K, a temperature previously difficult to access with scanning SQUID microscopy.

\begin{figure}
    \includegraphics[width=\linewidth]{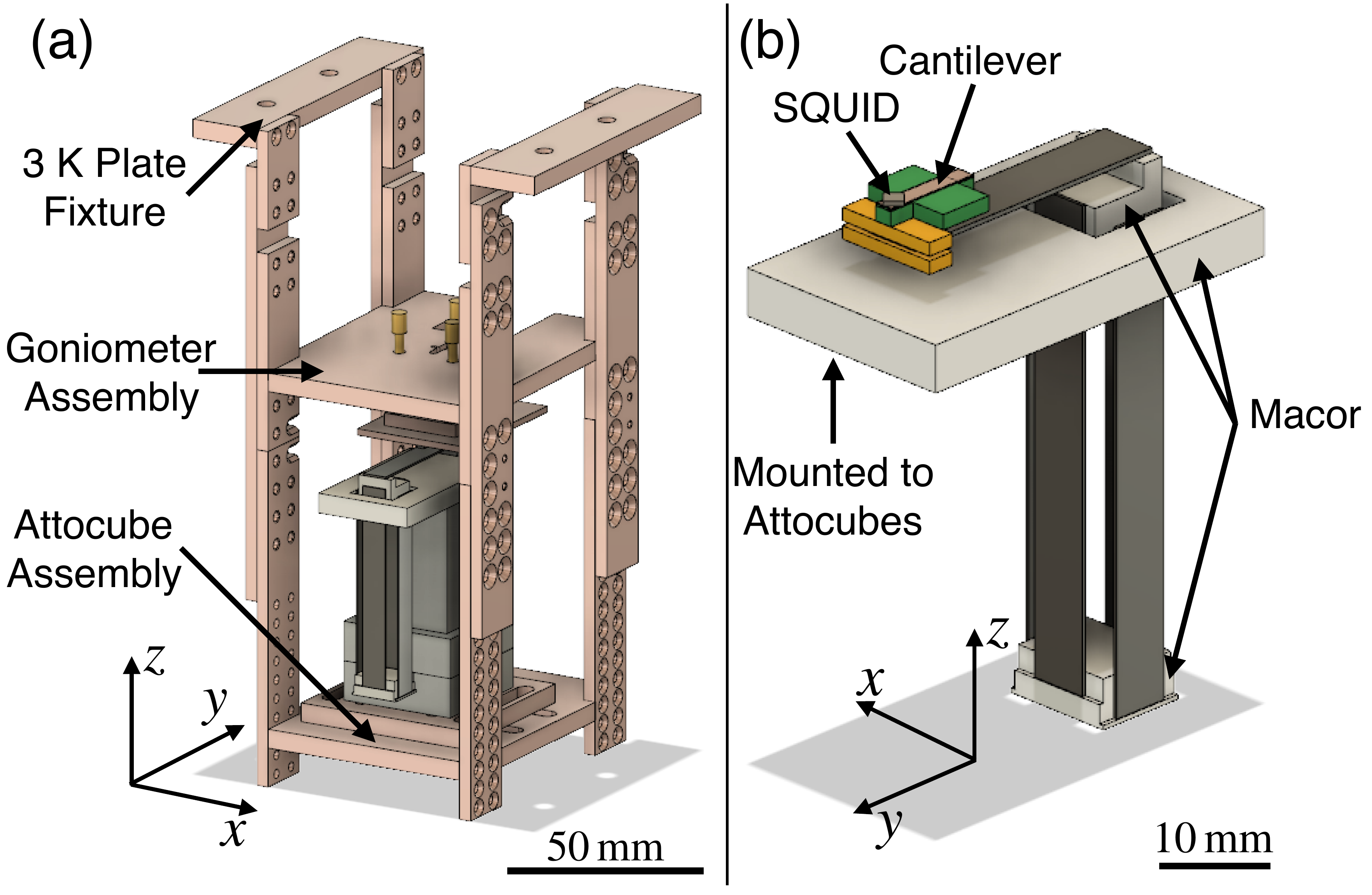}
    \caption{\label{fig:cage_scanner}(a) CAD model of fully assembled cage. (b) CAD model of the piezoelectric scanner, with the SQUID sensor at the end of the $z$-axis piezo bender. The large Macor section of the scanner is mounted rigidly to the Attocube stack for coarse positioning.}
\end{figure}

\section{Microscope Design}
The microscope consists of an oxygen-free high thermal conductivity (OFHC) copper ``cage'' (Fig.~\ref{fig:cage_scanner}(a)) mechanically and thermally anchored to the 3 K plate of a Bluefors LD-4K cryostat, on which three Attocube coarse positioners and a long-range piezoelectric scanner are mounted. The cage has a modular design, with three main sections: 3 K plate fixture, sample goniometer assembly, and Attocube assembly. The sample goniometer and Attocube assemblies can be independently removed from and attached to the 3 K plate fixture, simplifying the procedure for swapping out the sample and/or sensor. This modular design allows for more convenient mounting of the sensor, sample, and cage (relative to a more compact, single-piece design) at the expense of mechanical rigidity.

The Attocube ``stack'' consists of two ANPx311 horizontal stepper positioners ($x$ and $y$), and one ANPz102 vertical stepper positioner ($z$), which together provide a coarse positioning range of $6\times6\times5\;\mathrm{mm}^3$. This large coarse positioning range allows for mounting and measurement of multiple samples in a single cooldown.

The piezoelectric scanner (Fig.~\ref{fig:cage_scanner}(b)), based on the design by Siegel, \emph{et al.},\cite{benders_rsi1995} is optimized for long-range scanning at cryogenic temperatures. It consists of five piezoelectric bending actuators (two each for the $x$ and $y$ axes, and one for the $z$ axis) epoxied to three Macor pieces in a geometry that allows independent long-range motion along all three axes. The dimensions of the $x$-$y$ ($z$) piezoelectric benders are $50.80\times6.35\times0.38\;\mathrm{mm}$ ($31.75\times6.35\times0.51\;\mathrm{mm}$).  At 3 K, the displacement coefficients for the three axes are $dx/dV=dy/dV=0.9\;\um/\mathrm{V}$ and $dz/dV=0.1\;\um/\mathrm{V}$. The scanner is driven by three $\pm10\;\mathrm{V}$ analog outputs of a multifunction DAQ (NI USB-6363), through a commercial high-voltage amplifier with $20\times$ gain (Attocube ANC250), resulting in a total scan range of roughly $350\times350\times40\;\um^3$. This scanner design optimizes for linearity and scan range, again at the expense of mechanical rigidity.

\begin{figure}
    \includegraphics[width=\linewidth]{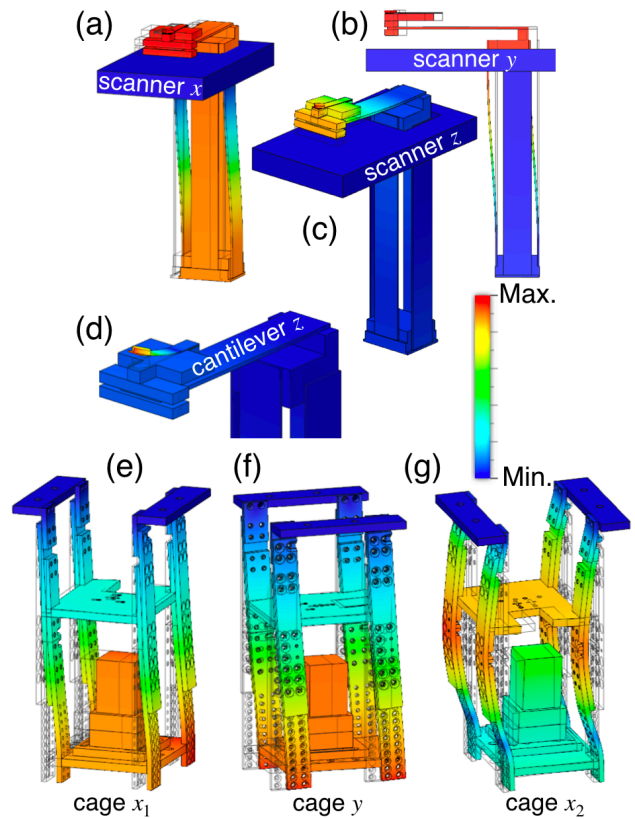}
    \caption{\label{fig:vib_modes}(a-d) Simulated mechanical modes of the scanner and cantilever, which we estimate to have resonant frequencies of 70 Hz, 100 Hz, 200 Hz, and 1 kHz, respectively. (e-g) Simulated modes of the cage, which we estimate to have resonant frequencies of 150 Hz, 250 Hz, and 650 Hz, respectively. Color scale indicates displacement normalized to the (arbitrary) maximum displacement of each mode. Simulations were performed using Autodesk Fusion 360 with room temperature material parameters to gain a qualitative understanding of the microscope's main vibration modes. Listed resonant frequencies are approximate; more precise values would require the models to include cryogenic material parameters and knowledge of the added mass from electrical leads, etc.}
\end{figure}

The SQUID sensor is mounted at the free end of a $\sim~2\times10\times0.12\;\mathrm{mm}^3$ Cu cantilever, which makes up half of a parallel plate capacitor on a printed circuit board (PCB) mounted at the end of the $z$-axis piezo bender (Fig.~\ref{fig:cage_scanner}(b)). Mechanical contact between the sensor and the sample is detected by monitoring the cantilever capacitance. The cantilever has a mechanical resonance (Fig.~\ref{fig:vib_modes}(d)) that we expect to contribute to relative motion between the sensor and sample in the $z$ direction near 1 kHz.

With a cage and scanner optimized for experimental flexibility and scan range over mechanical rigidity, the microscope is susceptible to excitation of mechanical resonances by periodic displacement of the 3 K plate due to the cryostat's pulse tube and other vibrations. In order to quantify the effect that such vibrations have on scanning SQUID measurements, we must measure the frequency spectrum of the relative displacement between sensor and sample along all three axes.

%
%
\section{Vibration Characterization}

\begin{figure}
    \includegraphics[width=0.9\linewidth]{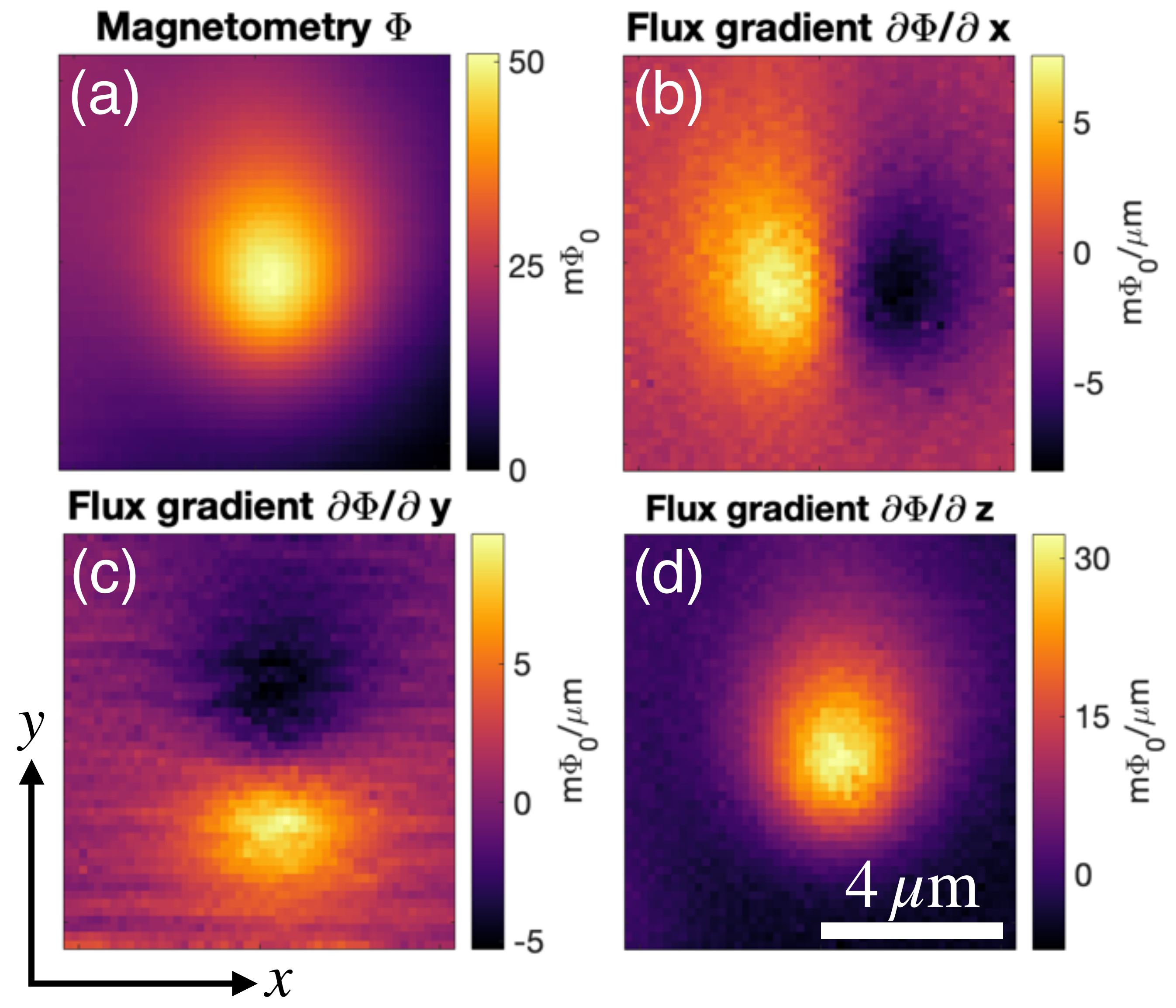}
    \caption{\label{fig:flux_gradients}(a) Magnetometry image of a vortex in a Nb film, measured with a $0.6\;\um$ inner diameter pickup loop.\cite{sssm_rsi2016} (b,c) Numerical gradients of the flux data from (a) in the $x$ and $y$ directions, respectively. (d) $z$-gradient of the flux, calculated from magnetometry maps at two heights above the Nb film, separated by $\Delta z=0.4\,\um$.}
\end{figure}

\begin{figure}
    \includegraphics[width=0.6\linewidth]{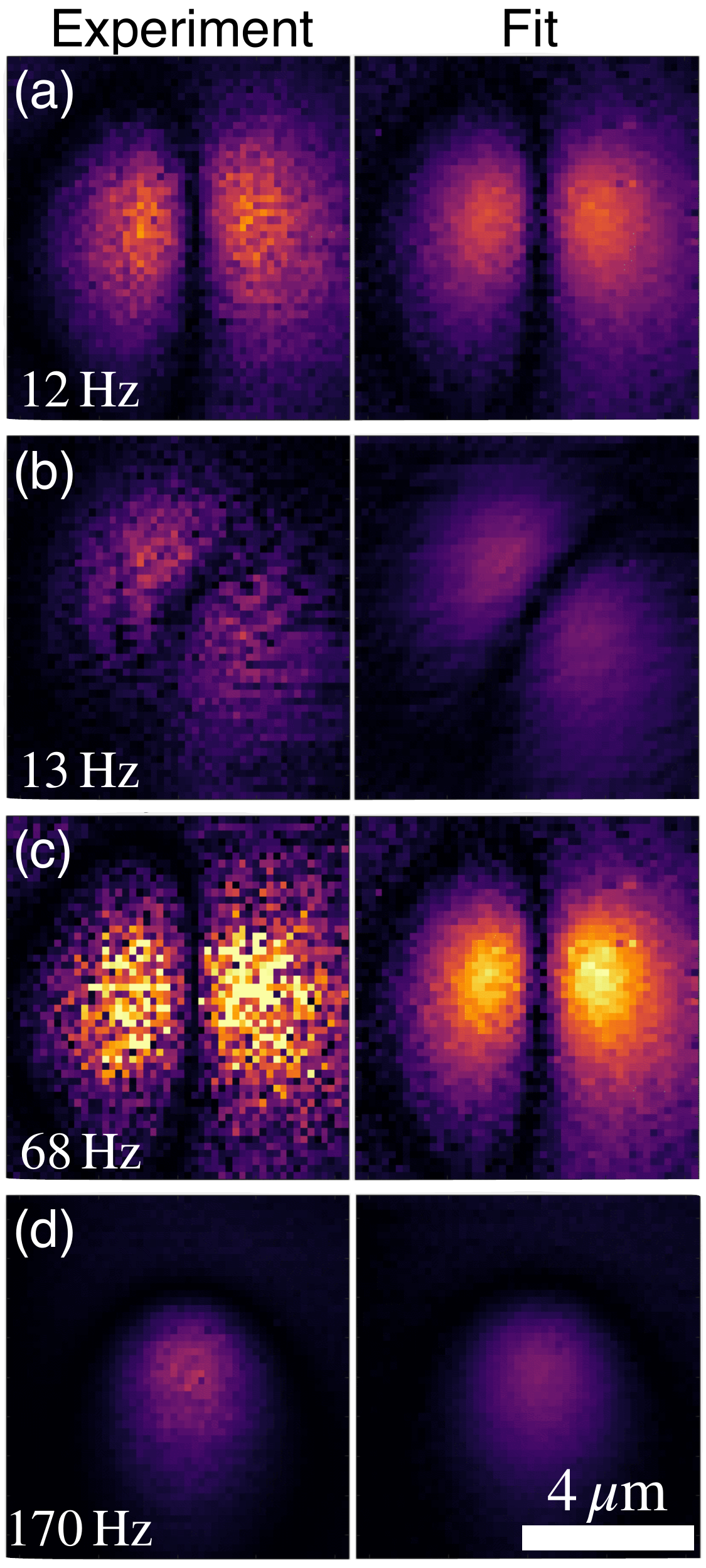}
    \caption{\label{fig:vib_fits}SQUID noise spectral density (arb. units) as a function of position near a vortex at (a) 12 Hz, (b) 13 Hz, (c) 68 Hz, and (d) 170 Hz, with fits to the model described in Ref.~\onlinecite{apl_vib2016}. Panels (a) and (c) correspond to vibrations primarily in the $x$ direction, panel (c) shows a combination of $x$ and $y$ vibrations, and panel (d) shows a higher-frequency $z$-dominated mode.}
\end{figure}

We characterize the frequency spectrum of relative sensor-sample vibrations by measuring the SQUID noise spectrum as a function of position in a region of sharp flux gradient.\cite{apl_vib2016} An isolated vortex in a Nb film has a spatial extent given by the London penetration depth $\lambda\approx80\;\mathrm{nm}$, well below the spatial resolution of the SQUID sensors used in this work (which have a pickup loop with an inner diameter of $600\;\mathrm{nm}$\cite{sssm_rsi2016}). As such, we can treat the vortex as a magnetic point source with sharp, experimentally measurable flux gradients in the $x$, $y$, and $z$ directions (Fig.~\ref{fig:flux_gradients}). Relative motion between the sensor and the sample at a given frequency is picked up as noise in the SQUID flux signal at that frequency, with spatial dependence determined by the direction of the motion, and the flux gradient in that direction (see Appendix~\ref{appendix} for further details).

Fig.~\ref{fig:vib_fits} shows exemplary maps of SQUID noise spectral density as a function of position near a vortex measured with the microscope mounted rigidly to the 3 K plate, along with fits to the model described in Ref.~\onlinecite{apl_vib2016} and Appendix~\ref{appendix}. Fig.~\ref{fig:vib_fits}(a) and \ref{fig:vib_fits}(c) show noise density at 12 Hz and 68 Hz respectively, indicating large amplitude sensor-sample motion primarily along the $x$ axis. In contrast, Fig.~\ref{fig:vib_fits}(b) indicates that the vibrations at 13 Hz are along both the $x$ and $y$ axes. Fig.~\ref{fig:vib_fits}(d) shows a higher-frequency mechanical mode dominated by motion along the $z$ axis. From these fits, we can extract estimates of the amplitude of the sensor-sample motion in all three directions as a function of frequency.

\begin{figure}
    \includegraphics[width=\linewidth]{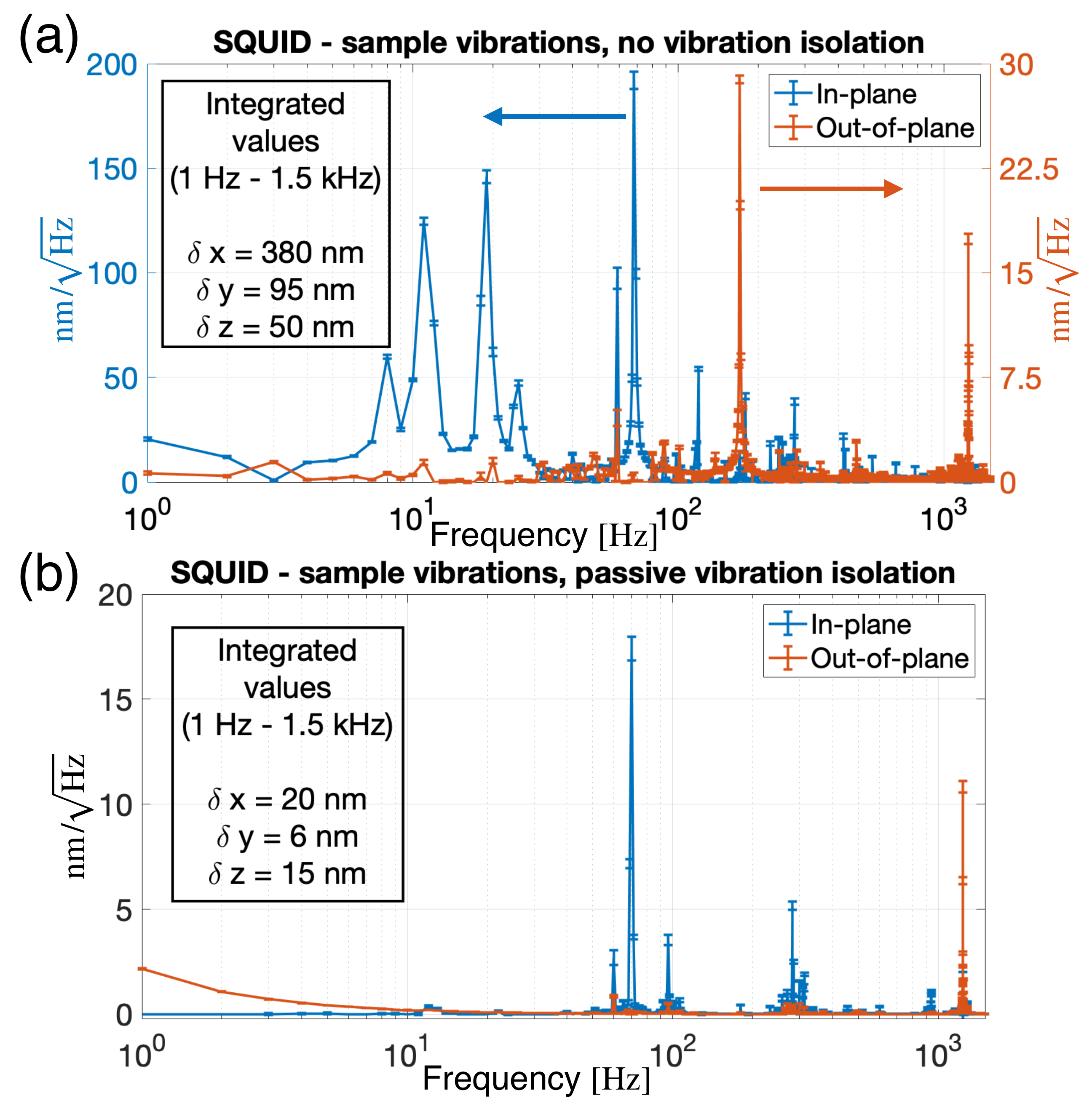}
    \caption{\label{fig:vib_spectrum}Sensor-sample vibration spectrum up to 1.5 kHz before (a) and after (b) implementation of passive vibration isolation, extracted from fits to SQUID noise near an isolated vortex (see Appendix~\ref{appendix}). Insets: Sensor-sample displacement along the $x$, $y$, and $z$ axes, integrated from 1 Hz to 1.5 kHz. Error bars are 95\% confidence intervals from least squares fitting.}
\end{figure}

As shown in Fig.~\ref{fig:vib_spectrum}(a), with no mechanical isolation between the microscope and the 3 K plate of the cryostat, there are several low-frequency mechanical resonances resulting in in-plane sensor-sample displacements of a few hundred nanometers, and out-of-plane displacements of a few tens of nanometers. As we identified no mechanical resonances in the scanner or cage below $\sim70\;\mathrm{Hz}$ (Fig.~\ref{fig:vib_modes}), we suspect that the broad low-frequency ($\sim6\;\mathrm{Hz}-30\;\mathrm{Hz}$) peaks in Fig.~\ref{fig:vib_spectrum}(a) are due to beating between scanner and cage modes.

Relative sensor-sample motion of this magnitude poses a problem for scanning SQUID measurements involving sharp flux gradients (such as imaging vortices in superconductors), measurements with strong dependence on the mutual inductance between the SQUID pickup loop and sample (such as measurements of ring-like samples), and measurements of strongly paramagnetic or diamagnetic materials.

\section{Passive Vibration Isolation}
The simplest way to isolate the microscope from motion of the 3 K plate is to insert a mechanical low-pass filter between the cage and the 3 K plate such that displacements of the 3 K plate at frequencies corresponding to resonances in the cage and scanner are attenuated rather than transferred to the microscope. A simple mass and spring system acts as mechanical low-pass filter with transfer function $G(\omega)=\omega_0^2/\sqrt{(\omega^2-\omega^2_0)^2+(\omega\omega_0/Q)^2}$, where $\omega$ is frequency of the excitation, $\omega_0=2\pi f_0$ is the resonant frequency of the filter, and $Q$ is the mechanical quality factor. In order to effectively isolate the microscope from the 3 K plate, we must design a filter with resonant frequency $\omega_0$ well below the frequency $\omega$ of any displacement of the 3 K plate and of any resonances of the microscope, such that $G(\omega)\ll1$.

To achieve this, we mount the 3 K plate fixture to the 3 K plate via four steel springs, each with an unextended length of $L_0=6.35\;\mathrm{cm}$ and nominal spring constant of $k=33.3\;\mathrm{N/m}$ at 300 K. The resonant frequency $\omega_0=2\pi f_0$ of a simple mass and spring system in Earth's gravitational field is given by $\omega_0=\sqrt{g/\Delta L}$, where $g=9.8\;\mathrm{m/s}^2$ and $\Delta L\equiv L_\text{extended}-L_0$ is the extension of the spring due to gravitational force on the mass. The steel springs are rated for a maximum extension of $\Delta L=12.29\;\mathrm{cm}$, giving a lower bound on the resonant frequency of the system of $f_0=1.42\;\mathrm{Hz}$. The actual extension of the springs when the microscope is suspended at 300 K is approximately $9\;\mathrm{cm}$, corresponding to a resonant frequency of $f_0=1.66\;\mathrm{Hz}$. The low temperature resonant frequency of the spring system is likely slightly higher due to the increased shear modulus of steel at low temperature. The cage remains thermally anchored to the 3 K plate with OFHC Cu ribbons, such that the cryostat, microscope, and sample can be cooled from 300 K to 3 K in 14 hours (cooldown time without vibration isolation was 12 hours).

The frequency spectrum of relative sensor-sample motion after implementation of the passive vibration isolation described above is shown in Fig.~\ref{fig:vib_spectrum}(b). The sensor-sample displacement amplitude (displacement spectral density integrated from $1\;\mathrm{Hz}$ to $1.5\;\mathrm{kHz}$) is reduced from roughly $400\;\mathrm{nm}$ to $20\;\mathrm{nm}$ in-plane, and from roughly $50\;\mathrm{nm}$ to $15\;\mathrm{nm}$ out-of-plane. Most of the remaining in-plane spectral weight is carried by the fundamental vibration mode of the scanner at $\sim70\;\mathrm{Hz}$ (Fig.~\ref{fig:vib_modes}(a)), while most of the remaining out-of-plane spectral weight is carried by what is likely the Cu cantilever's fundamental vibration mode at $\sim1240\;\mathrm{Hz}$ (Fig.~\ref{fig:vib_modes}(d)).
%
%
\section{Variable temperature sample stage}
Variable sample temperature SQUID microscopes have been developed with high-$T_\mathrm{c}$\cite{tzalenchuk1999variable,fleet2001closed} and low-$T_\mathrm{c}$\cite{variable_apl1999,baudenbacher2002high,morooka2000development} sensors, and with the sensor and sample in the same vacuum space\cite{variable_apl1999,morooka2000development,tzalenchuk1999variable} or with the SQUID separated from the sample by a thin vacuum barrier.\cite{fleet2001closed,baudenbacher2002high}
Our scanning SQUID microscopes, which use low-$T_\mathrm{c}$ SQUID susceptometers within one micron of contact with the sample for the highest sensitivity and spatial resolution, require that the Nb SQUID sensor and Nb series SQUID array amplifier\cite{array_ieee2001} be cooled well below their critical temperature of $9.2\;\mathrm{K}$. Although there is typically not direct mechanical contact between the SQUID sensor and the sample during scanning, there is inevitably some thermal coupling between the two, which in practice has limited the sample temperature range available to scanning SQUID to roughly 100 K.

Wet variable sample temperature SQUID microscopes with Nb SQUID sensors require an exchange gas while cooling, and careful tuning of the amount of this exchange gas to attain high sample temperatures while keeping the SQUID below its transition temperature.\cite{variable_apl1999} In contrast, our dry system cools without an exchange gas, and we can vary the sample temperature from 3 K to 110 K continuously while keeping the SQUID superconducting. 

\begin{figure}
    \includegraphics[width=\linewidth]{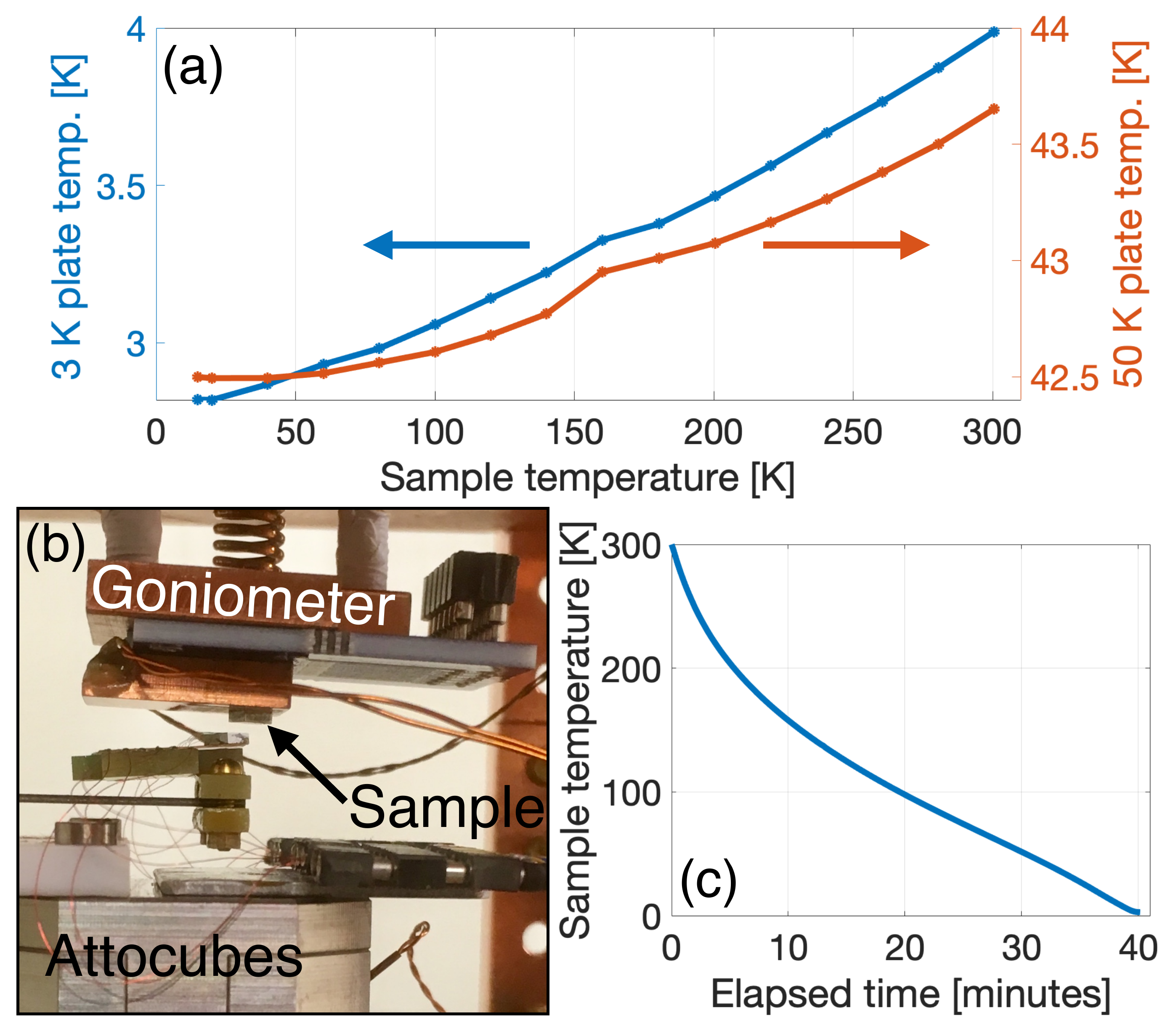}
    \caption{\label{fig:temps}(a) Temperature of the 3 K and 50 K plates of the cryostat as a function of sample temperature. (b) Photograph of the variable-temperature sample stage mounted on the goniometer. (c) Sample temperature vs. time after heating the sample to 300 K then turning off the sample heater.}
\end{figure}

The large experimental volume available in the Bluefors LD-4K cryostat and the open modular cage design described above allow for careful thermal isolation of the sample from the rest of the microscope. Fig.~\ref{fig:temps}(b) shows the goniometer with variable temperature sample mount attached, with the SQUID sensor and Attocube stack below. The goniometer is used to adjust the alignment between the SQUID sensor and sample at room temperature.

The sample mount consists of a $\sim10\times10\times2\;\mathrm{mm}$ OFHC Cu block onto which a Si diode thermometer and resistive heater are attached with EPO-TEK H70E thermally conductive epoxy. The thermometer leads are made of narrow gauge, low thermal conductivity phosphor bronze, so nearly all of the cooling power for the sample mount comes from the two heater leads, which are 28 AWG Cu magnet wire.

The heater and thermometer leads, as well as any leads for electrical contact to the sample, are insulated by vacuum ($<1\times10^{-6}$ Torr) from the microscope and all other electrical leads, and thermally anchored to a Cu bobbin attached rigidly to the 3 K plate. Thus the only thermal link between the sample mount and the microscope is through the $1.6\;\mathrm{mm}$ thick FR-4 PCB substrate separating the sample mount from the goniometer, which has a thermal conductivity of approximately $0.05\;\mathrm{W/m}\cdot\mathrm{K}$ at 3 K.\cite{fr4_therm2009} With the cryostat at base temperature, the cooling power of the heater leads is sufficient to cool the sample from 300 K to 3 K in roughly 40 minutes (Fig.~\ref{fig:temps}(c)).
\begin{figure}
    \includegraphics[width=0.8\linewidth]{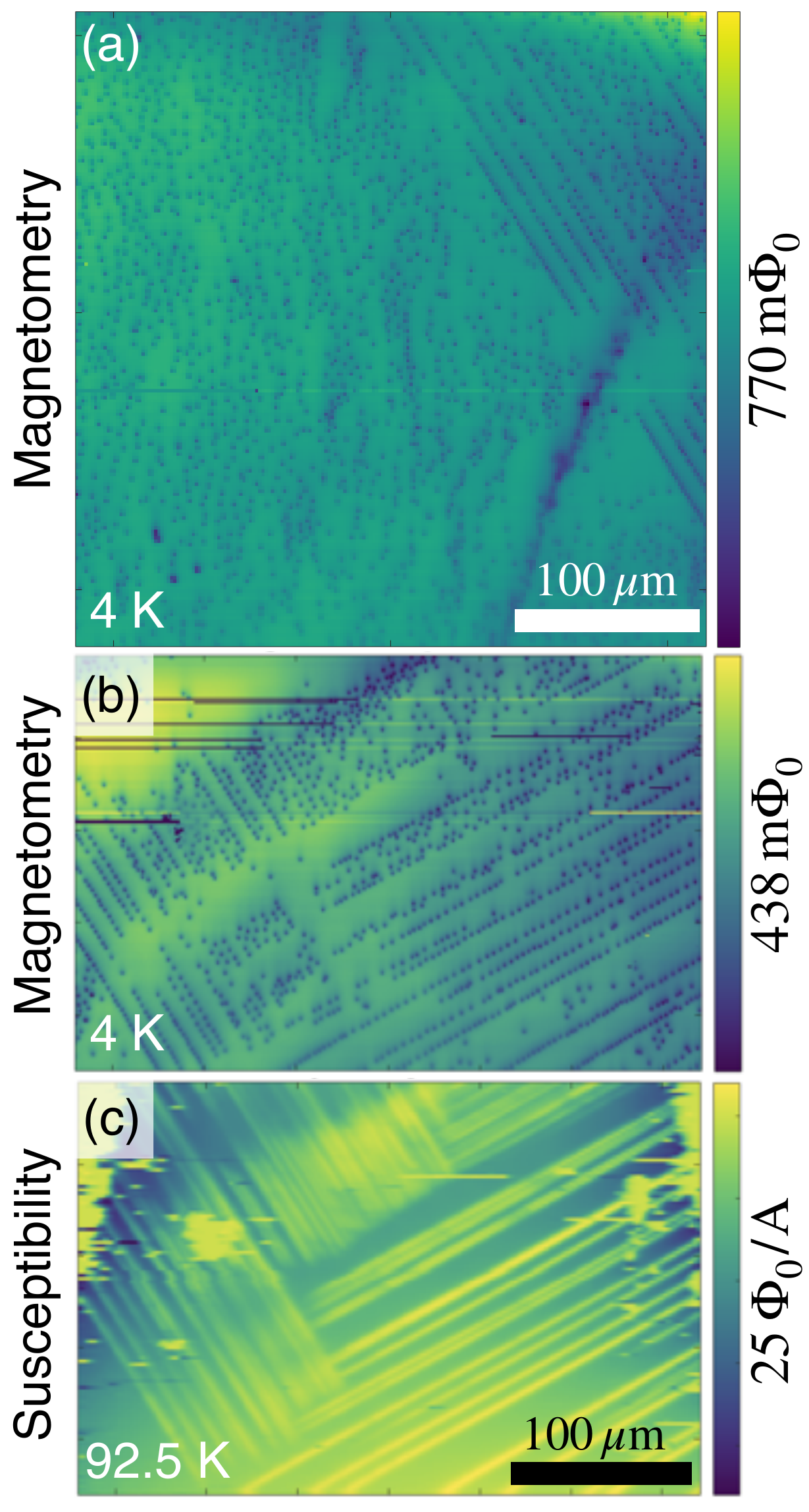}
    \caption{\label{fig:big_scanYBCO}(a) Magnetometry map of inhomogeneous vortex pinning in optimally-doped single crystal YBCO cooled in an applied field, demonstrating the microscope's $\sim350\times350\;\um^2$ scan range. (b, c) Magnetometry (b, 4 K) and susceptibility (c, 92.5 K) maps of the same region on the sample, showing correspondence between vortex pinning and reduced diamagnetic response on twin domain boundaries. The full-scale diamagnetic response is roughly $70\;\Phi_0/\mathrm{A}$ (darker $\leftrightarrow$ more diamagnetic). Horizontal line features are scanning artifacts due to the SQUID flux-locked loop unlocking momentarily.}
\end{figure}
\begin{figure*}
    \includegraphics[width=\linewidth]{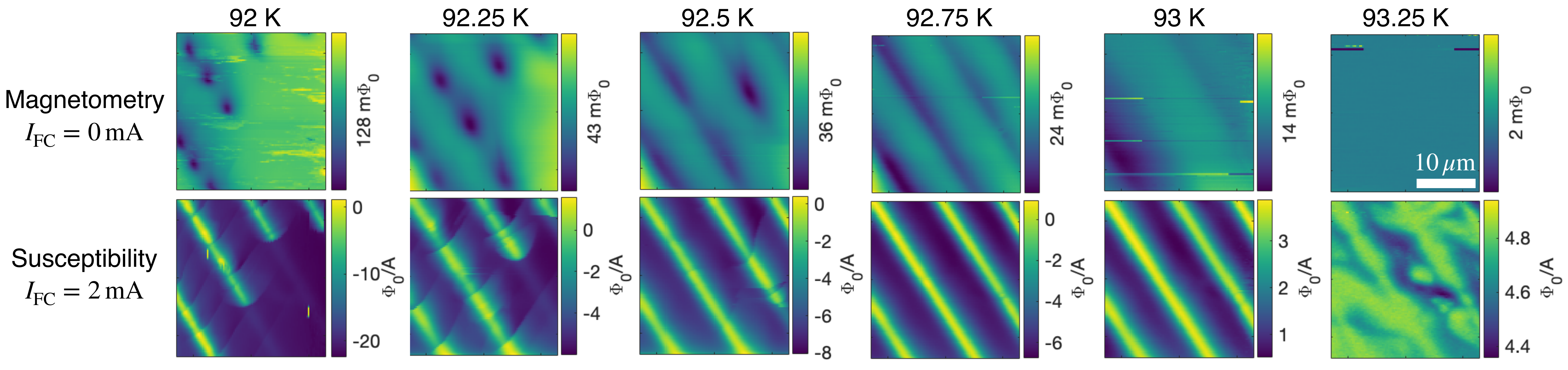}
    \caption{\label{fig:vortices_tc}Magnetometry (top row, no field coil current) and susceptibility (bottom row, 2 mA  rms field coil current) maps of a twinned region of the YBCO sample as a function of temperature through $T_\mathrm{c}$. Vortices pinned on twin domain boundaries spread out as the London penetration depth $\lambda$ diverges near $T_\text{c}$. Sharp arc-like features in susceptibility below 92.75 K are from vortices moving along the twin boundaries due to the Lorentz force from the ac field coil current. These features disappear when the flux is no longer strongly localized in vortices.}
\end{figure*}

This balance between cooling of the sample and isolation of the sample mount from the microscope allows us to measure samples at temperatures from 2.8 K to roughly 110 K with nearly constant flux sensitivity over that range. To demonstrate, we map the diamagnetic susceptibility of a high quality optimally-doped YBCO single crystal as a function of temperature near $T_\text{c}\approx 93\;\mathrm{K}$ (Fig.~\ref{fig:big_scanYBCO}(b,c), Fig.~\ref{fig:vortices_tc}). We find that the diamagnetic response is strongly suppressed near $T_\text{c}$ along twin domain boundaries, and that, unsurprisingly, vortices preferentially pin in these regions of suppressed diamagnetism when the sample is cooled through $T_\text{c}$ in a magnetic field (Fig.~\ref{fig:big_scanYBCO}(c)). (The uniformly-colored regions in the upper corners of the susceptibility map in Fig.~\ref{fig:big_scanYBCO}(c) are due to the SQUID sensor briefly coming into contact with debris on the sample surface.)

At sample temperatures above 110 K, the critical currents of both the SQUID sensor and series array amplifier are diminished significantly, which reduces signal-to-noise and indicates that the entire microscope has been heated to nearly 9 K. While scanning SQUID operation is currently limited to 110 K, we can in fact heat the sample mount to 300 K while only increasing the cryostat's 3 K plate temperature to 4 K (Fig.~\ref{fig:temps}(a)). This suggests there is a temperature gradient of several Kelvin along the Cu ribbons that thermally anchor the cage to the 3 K plate.

There is no indication that black body radiation from the sample mount is measurably heating the SQUID sensor, as the heating has little dependence on sensor-sample distance. The limiting factor appears to be conduction through the FR-4 PCB substrate, whose thermal conductivity increases by roughly an order of magnitude from 3 K to 100 K.\cite{fr4_therm2009} This upper limit on sample temperature could be increased by reducing the effective surface area or increasing the thickness of the insulating layer between the sample mount and goniometer.

%
%
\section{Conclusion}
In conclusion, we have characterized sensor-sample vibrations in a cryogen-free scanning SQUID microscope and implemented passive vibration isolation to reduce these vibrations below our threshold for detection over most of the frequency spectrum. We then created a variable temperature sample stage, enabling measurement of samples at temperature from 2.8 K to 110 K, and demonstrated these capabilities by mapping inhomogeneous diamagnetic susceptibility and vortex pinning in optimally-doped YBCO above 90 K. Together with sub-micron spatial resolution, few-$\mu\Phi_0/\sqrt{\mathrm{Hz}}$ white noise floor,\cite{sssm_rsi2016} and $350\times350\;\um^2$ scan range (Fig.~\ref{fig:big_scanYBCO}(a)), these advances position us for further studies of superconducting and magnetic materials and devices over a temperature range previously difficult to access with scanning SQUID microscopy.

This microscope is compatible with our dc scanning SQUID susceptometers,\cite{sssm_rsi2016} scanning SQUID susceptometers with dispersive readout,\cite{dispersive_apl2018} and scanning SQUID samplers.\cite{sampler_rsi2017} We plan to use this microscope for studies of unconventional superconductors, viscous electron flow in condensed matter systems, dc and time-resolved studies of Josephson effects, and time-resolved studies of vortex dynamics.

\begin{acknowledgments}
We acknowledge David Low for useful discussions, and Doug Bonn and Ruixing Liang for providing the YBCO sample. This work is supported by the Department of Energy, Office of Science, Basic Energy Sciences, Materials Sciences and Engineering Division, under Contract DE-AC02-76SF00515. The SQUID susceptometers used in this work were developed under an NSF IMR-MIP Grant No. DMR-0957616.
\end{acknowledgments}

\appendix
\section{\label{appendix}Vibration Characterization Method}
We measure the SQUID flux noise spectrum $\tilde{\Phi}(x,y,z,f)$ at a given pickup loop position $(x,y,z)$ by taking the Fourier transform of the flux signal $\Phi(x,y,z,t)$ acquired for 1 s at a sampling rate of 1 MHz. The zero-frequency component of $\Phi(x,y,z,f)$ measured at a constant height $z=z_0$ above the sample is our dc magnetometry data $\Phi(x,y,z_0)=\langle{\Phi(x,y,z_0,t)\rangle}_t$ , where $\langle{\rangle}_t$ denotes the average over the aquisition time (Fig.~\ref{fig:flux_gradients}(a)). We numerically calculate the gradients of $\Phi(x,y,z_0)$ in the $x$ and $y$ directions (Fig.~\ref{fig:flux_gradients}(b,c)), and approximate the gradient in the $z$ direction (Fig.~\ref{fig:flux_gradients}(d)) by subtracting magnetometry data taken at two different heights above the sample, separated by $\Delta z=0.4\,\um$:
$$\left.\frac{\partial\Phi(x,y,z)}{\partial z}\right|_{z=z_0}\approx\frac{\Phi(x,y,z_0) - \Phi(x,y,z_0-\Delta z)}{\Delta z}.$$

Assuming a stationary vortex located at the origin, we can approximate the time-dependent flux signal $\Phi(x,y,z_0,t)$ at height $z_0$ above the sample to first order as
$$\Phi(x,y,z_0,t)=\Phi(0,0,z_0)+\vec{\nabla}\left.\Phi(x,y,z)\right|_{z=z_0}\cdot\vec{r}(t)+\eta(t),$$
where $\Phi(0,0,z_0)$ is the actual flux at height $z_0$ above the center of the vortex, $\vec{\nabla}=\left(\frac{\partial}{\partial x}\hat{\vec{x}}+\frac{\partial}{\partial y}\hat{\vec{y}}+\frac{\partial}{\partial z}\hat{\vec{z}}\right)$, $\vec{r}(t)=x(t)\hat{\vec{x}}+y(t)\hat{\vec{y}}+z(t)\hat{\vec{z}}$ is the time-dependent position of the SQUID pickup loop relative to the center of the vortex, and $\eta(t)$ is any position-independent noise on the flux signal.

Fourier transforming the above expression, we find
\begin{equation}\label{eq:fit}|\tilde{\Phi}(x,y,z_0,f)|=\left|\vec{\nabla}\left.\Phi(x,y,z)\right|_{z=z_0}\cdot\tilde{\vec{r}}(f)+\tilde{\eta}(f)\right|,\end{equation}
where $\tilde{\vec{r}}(f)=\tilde{x}(f)\hat{\vec{x}}+\tilde{y}(f)\hat{\vec{y}}+\tilde{z}(f)\hat{\vec{z}}$ is a vector of the SQUID-sample displacement spectral density, and $\tilde{\eta}(f)$ is the spectral density of position-independent flux noise. Due to the cylindrical symmetry of the vortex, it is convenient to work in cylindrical coordinates, where $\tilde{\vec{r}}(f)=\tilde{\rho}(f)(\cos\tilde{\theta}(f)\hat{\vec{x}}+\sin\tilde{\theta}(f)\hat{\vec{y}})+\tilde{z}(f)\hat{\vec{z}}$, with $\tilde{\rho}\geq0$ and $0\leq\tilde{\theta}<2\pi$.

After measuring the flux noise spectrum on a grid of points ($x$,$y$) at height $z_0$ near an isolated vortex and calculating the flux gradients $\vec{\nabla}\left.\Phi(x,y,z)\right|_{z=z_0}$ as described above, for each frequency $f$ up to $f_\text{max}=1.5\,\mathrm{kHz}$ we perform a least squares fit of the measured $|\tilde{\Phi}(x,y,z_0,f)|$ to Equation~\ref{eq:fit} with $\tilde{\rho}$, $\tilde{\theta}$, $\tilde{z}$, and $\tilde{\eta}$ as free parameters. The best-fit values for $\tilde{\rho}(f)$ and $\tilde{z}(f)$ are shown in Fig.~\ref{fig:vib_spectrum}, labeled as ``in-plane'' and ``out-of-plane,'' respectively.

The sensitivity of this method is determined by the flux sensitivity of the SQUID sensor and the magnitude of the flux gradient in which the sensor moves. Our sensors have a typical white noise floor of a few $\mu\Phi_0/\sqrt{\mathrm{Hz}}$ with a $1/f$ tail below $\sim50\,\mathrm{Hz}$.\cite{sssm_rsi2016} For the sensor used in this work, which has a pickup loop with inner diameter $0.6\,\um$, the magnitude of the in-plane flux gradient near an isolated vortex is approximately $5\,\mathrm{m}\Phi_0/\um$ (Fig.~\ref{fig:flux_gradients}(b,c)). Assuming a $1\,\mu\Phi_0/\sqrt{\mathrm{Hz}}$ flux noise floor, this gives a sensor-sample displacement noise floor for in-plane vibrations of approximately $(1\,\mu\Phi_0/\sqrt{\mathrm{Hz}})/(5\,\mathrm{m}\Phi_0/\um)=0.2\,\mathrm{nm}/\sqrt{\mathrm{Hz}},$ with a slightly lower noise floor for out-of-plane sensor-sample displacement due to the steeper flux gradient in the $z$ direction (Fig.~\ref{fig:flux_gradients}(d)). However, sensitivity to low-frequency sensor-sample displacement is reduced by the $1/f$ tail in the flux sensitivity. Therefore we estimate the low-frequency sensitivity of this method to be of order $1\,\mathrm{nm}/\sqrt{\mathrm{Hz}}.$

\bibliography{bf4k_rsi}
\end{document}